# Circular polarization in comets: Observations of comet C/1999 S4 (LINEAR) and tentative interpretation

Vera Rosenbush[a,*], Ludmilla Kolokolova[b], Alexander Lazarian[c], Nikolai Shakhovskoy[d], Nikolai Kiselev[a]

[a] *Main Astronomical Observatory of National Academy of Sciences, Zabolotnoho str. 27, 03680 Kyiv, Ukraine*
Corresponding Author E-mail address: rosevera@mao.kiev.ua
[b] *University of Maryland, College Park, MD 20740, USA*
[c] *University of Wisconsin, Madison, WI 53706, USA*
[d] *Crimean Astrophysical Observatory, Nauchny, Ukraine*

Pages:    47
Tables:   2
Figures:  5

**Proposed Running Head:** Circular polarization in comet C/1999 S4 (LINEAR)

**Editorial correspondence to:**

Dr. Vera Rosenbush
Main Astronomical Observatory
of National Academy of Sciences
Zabolotnoho str. 27,
03680 Kyiv,
UKRAINE

Phone:      380 (44) 526–21–47
Fax:        380 (44) 526–21–47
E-mail address: rosevera@mao.kiev.ua

**ABSTRACT**

Comet C/1999 S4 (LINEAR) was exceptional in many respects. Its nucleus underwent multiple fragmentations culminating in the complete disruption around July 20, 2000. We present circular polarization measurements along the cuts through the coma and nucleus of the comet during three separate observing runs, in June 28-July 2, July 8-9, and July 21-22, 2000. The circular polarization was detected at a rather high level, up to 0.8%. The left-handed as well as right-handed polarization was observed over the coma with the left circularly polarized light systematically observed in the sunward part of the coma. During our observations the phase angle of the comet varied from 61 up to 122°, which allowed us to reveal variations of circular polarization with the phase angle. Correlation between the degree of circular polarization, visual magnitude, water production rate, and linear polarization of comet C/1999 S4 (LINEAR) during its final fragmentation in July 2000 was found. The mechanisms that may produce circular polarization in comets and specifically in comet C/1999 S4 (LINEAR) are discussed and some tentative interpretation is presented.

# 1. Introduction

After successful exploration of the properties of linear polarization of cometary dust there arises interest about its circular polarization. Dozens of previously published papers show that circular polarization is sensitive to the shape, structure and composition of the scatterers and, thus, may provide further proof of the complex structure of cometary grains and put constraints on their shape and composition. However, measurements of circular polarization (hereafter CP) are still rare and conditions for its formation in comets are poorly understood. Bandermann and Kemp (1973) and Dolginov and Mytrophanov (1975) estimated that the highest possible degree of circular polarization in comets, arising from scattering on icy and silicate aligned grains, can be rather significant, reaching 2-4%. However, attempts to detect circular polarization in comets C/1969 T1 (Tago−Sato−Kosaka), C/1973 E1 (Kohoutek), C/1974 C1 (Bradfield), and C/1975 VI (West) were unsuccessful (Wolf, 1972; Michalsky, 1981).

Comet 1P/Halley was the first comet for which circular polarization was reliably detected. Left-handed (negative) circular polarization was observed on March 16-20, 1986 (Metz and Haefner, 1987), but on April 8-28 it had variable sign (Dollfus and Suchail, 1987; Morozhenko et al., 1987). The observed values of CP were highly variable over the coma and showed temporal variations, from 0.04% up to 2.3%. The degree of circular polarization was found to be strongly dependent on the aperture size. Apparently, prior to comet Halley attempts to detect circular polarization were unsuccessful because of the use of large diaphragms, not to mention that all of these observations, including observations of comet Halley, were characterized by a low accuracy.

The first high-precision measurements of CP over the coma, including the nucleus, shells, regions between shells, and tail, were obtained for comet Hale-Bopp by Rosenbush et al. (1997). Only left circularly polarized light with a maximum value −0.26 ± 0.02% was detected for all measured areas in the coma on March 11, 1997. The polarization was close to zero (less than 0.08%) at the cometary nucleus. Spatial variations up to 0.1% in the degree of CP were observed. Manset and Bastien (2000) also measured CP in this comet. The absolute value was found at the same level, ≤ 0.3%, both left and right circularly polarized light being detected during April 2-16, 1997.

Comet C/1999 S4 (LINEAR), hereafter called S4, was discovered on September 27, 1999 by the Lincoln Near Earth Asteroid Research (LINEAR) program at heliocentric distance $r$ = 4.3 AU. The ephemeris predicted that the comet would be as close to the Earth as $\Delta$=0.364 AU just a few days before perihelion on 26 July 2000 ($r$ = 0.765 AU). Therefore the comet was expected to be observed at large phase angles, together with favorable brightness predictions. These conditions determined the main goal of our observations: to obtain the phase-angle dependence of CP of the light scattered by the dust in cometary atmosphere. Exceeding our expectations, during our observations from June 28 to July 22, 2000, comet S4 was in a very active state. Its nucleus underwent multiple fragmentations culminating in complete disruption around July 20. Thus, we got a unique possibility to study the outflow behavior of dust from the nucleus of disintegrating comet.

## 2. Observations and data reduction

*2.1. Instrument and procedure for measuring of circular polarization*

Measurements of CP in comet S4 were made with a one-channel photoelectric photometer-polarimeter (Shakhovskoy et al., 1998) mounted at the Cassegrain focus of the 2.6-m (f/16) Shain telescope of the Crimean Astrophysical Observatory. The photopolarimeter consists of a motionless polaroid, which is installed immediately behind an achromatic quarter-wave phase plate rotating with a frequency $\approx$33 Hz. Four counters are used for registration of photoelectrons (impulses). During one revolution of the quarter-wave phase plate, each counter switches on twice (for a time equal to one-quarter revolution), and the time when one pair of counters is open is shifted by one-half period with respect to the other pair. In such a manner each pair of counters produces four brightness measurements over a revolution period of the phase plate. The advantage of this instrument is its high efficiency in observations of faint objects and the fact that the results are independent of the light flux variations caused by the object variability as well as atmospheric transparency variations and inaccurate guiding.

The total number of impulses $N_i$ (where $i = 1, 2, 3, 4$), recorded by each counter, is determined by the following equations:

$$\begin{aligned}
N_1 &= C(1 + \frac{Q}{2} + \frac{2V}{\pi}\cos 2\psi_0), \\
N_2 &= C(1 + \frac{Q}{2} - \frac{2V}{\pi}\cos 2\psi_0), \\
N_3 &= C(1 + \frac{Q}{2} - \frac{2V}{\pi}\sin 2\psi_0), \\
N_4 &= C(1 + \frac{Q}{2} + \frac{2V}{\pi}\sin 2\psi_0),
\end{aligned} \quad (1)$$

where $Q$ and $V$ are the normalized Stokes parameters, $\psi_0$ is the angle which is determined by the position of the principal optical axis of the phase plate with respect to the principal plane of the analyzer, and $C$ is a coefficient dependent on the flux and transmission of the recording system. Using the fact that $N_1 + N_2 = N_3 + N_4 = 2C(1 + Q/2)$, one can define the following parameters:

$$\begin{aligned}
r_1 &= \frac{N_1 - N_2}{N_1 + N_2} = \frac{2V \cos 2\psi_0}{\pi(1 + Q/2)}, \\
r_2 &= \frac{N_3 - N_4}{N_3 + N_4} = \frac{2V \sin 2\psi_0}{\pi(1 + Q/2)}, \\
P_1 &= \frac{\pi}{2} r_1 (1 + Q/2) = V \cos 2\psi_0, \\
P_2 &= \frac{\pi}{2} r_2 (1 + Q/2) = V \sin 2\psi_0,
\end{aligned} \quad (2)$$

According to Shakhovskoy et al. (2001), the degree of circular polarization can be found from the following expression:

$$P_c = P_1 \cos 2\psi_0 + P_2 \sin 2\psi_0. \quad (3)$$

The angle $\psi_0$ is determined for every observation set from $P_1$ and $P_2$ by the orthogonal regression procedure (Isobe et al., 1990) and is equal to ~80°. The sign of circular polarization is determined by observing circularly polarized stars. By definition, circular polarization is right-handed ($+P_c$, positive) when the electric vector rotates clockwise, as viewed by an observer looking in the direction of the light propagation whereas the left-handed ($-P_c$, negative) polarization corresponds to the anticlockwise rotation of the electric vector (Mishchenko et al., 2002).

*2.2. Observations*

The comet was observed during three separate observing runs, in June 28-July 2, July 8-9, and July 21-22, 2000. The wide-band R filter and a 15″ diaphragm were used. The phase angle of the comet at the time of observations varied from 61 to 122°. The geocentric distance of the comet changed from 1.101 to 0.364 AU, while the heliocentric distance changed from 0.938 to 0.769 AU.

Every night the coma morphology was watched visually through the previewer of the photopolarimeter at the 2.6-m telescope. On June 28 the comet had the coma with well-defined central condensation and sharply outlined tail. On June 29 some very weak condensations was observed in the coma. The nucleus looked to be a double one and the sunward component of the nucleus was measured as brighter. However, on July 2 the single star-like nucleus was observed. On July 8 we again observed a weak condensation near the nucleus in the direction of the tail, and on July 21 and 22 two condensations of approximately equal brightness were clearly seen instead of a single nucleus. The brightness profiles through the nucleus have confirmed the presence of nuclear fragments that we visually observed. Note that on July 7 a fragment located ~460 km in projected distance tailward from the main nucleus was detected by the Hubble Space Telescope (Weaver et al., 2001).

To compensate the large proper motion of the comet, we used the following method of observations. The measurements of the total intensity and parameters determining the CP of the scattered light were made along cuts over the coma. The diaphragm was placed on the sunward side of the coma in such a way that each cut passed through the nucleus. The sections over the coma were determined by proper motion of the comet. The single exposure time on the coma was short and varied from 5 to 10 s depending on the proper motion of the comet. Such short integrations were needed because the comet could have drifted partially out of the aperture of the instrument. For all nights, except June 28, the CP of the dust radiation was measured along the cuts through the coma and nucleus. On June 28 the cut was made at the tailward side of the coma. To improve the signal-to-noise ratio, we obtained from 10 to 38 cuts during a given observational night, which were summarized subsequently. To minimize errors in the positioning, i.e., to be sure that the cometary nucleus always passed through the center of the diaphragm at different scans, we used cross hairs in the previewer and the guide.

Most of the comet observations were carried out in twilight or low above the horizon, i.e. through a high air mass. During all our observations good photometrical conditions prevailed except on July 9 and 22, when light cirri were sometimes in the sky. Since the sky background was varying non-linearly with time, it was measured after every 2-4 measurements of the comet, at the beginning and at the end of the cuts.

Several bright polarimetric standards with zero circular polarization were regularly measured to obtain the mean parameters of instrumental polarization $\overline{P}_1$ and $\overline{P}_2$ (Serkowski, 1974). The sign of circular polarization was determined by observing a circularly polarized star AM Her (Shakhovskoy et al., 2001).

Information about the observations is given in Table 1. In addition to heliocentric $r$ and geocentric $\Delta$ distances and phase angle $\alpha$, the position angle of comet radius-vector $PA$, the position angle of the cut $PA_{cut}$, and the diaphragm radius $\rho$ at the distance of the comet are listed. Because of the absence of an offset guide probe, the comet shifted in the diaphragm during one exposure by about $\Delta\rho$, which is also given in the table for each date. Thus, the real size of the diaphragm projected at the comet in the direction of the proper motion of the comet was equal to $2\rho + \Delta\rho$.

[**Table 1**]

*2.3. Data processing*

Initial data processing included reduction for instrumental efficiency (inequality of the operation time of counters), sky background, and instrumental polarization. The calculations of CP were carried out according to a special procedure, described by Shakhovskoy et al. (1998). The sky background radiation was approximated by a polynomial and removed during data reduction. Changes in intensity of the comet as well as the sky background were analyzed and all large deviations were eliminated from the data. The appearance of faint stars in the diaphragm could be the reason of these variations.

Taking into account the proper motion of the comet, we reduced the measurements of polarization and total intensity to corresponding distance from nucleus. The position of the maximum of the total intensity profile for each cut was used for superposition of the CP cuts obtained during the night; then the values of the CP at each point were averaged. Table 2 shows the night-averaged absolute values of circular polarization $P_c$. The data are presented as follows: date (UT) of the middle of the observations, phase angle $\alpha$, the mean overnight value of the cut-averaged circular polarization and its standard error $\sigma_{P_c}$, and the number of measurements along the cut. To estimate the total errors arising in our measurements of circular polarization, we investigated all possible sources of random and systematic errors.

[**Table 2**]

*2.4. Analysis of errors*

*2.4.1. Random errors*

The formulae (1-3) for determination of circular polarization are valid for the ideal case when the phase shift of the plate is exactly $\pi/2$ over the entire spectral range and each counter is in operation for a time equal to one-fourth of the plate's revolution period. In actual practice these conditions are fulfilled only approximately. In this case the random error is equal to:

$$\sigma_{P_c} = (\sigma_{P_1}^2 \cos^2 2\psi_0 + \sigma_{P_2}^2 \sin^2 2\psi_0)^{1/2} . \qquad (4)$$

If the errors of parameters $P_1$ and $P_2$ are only caused by the Poisson statistics of the measured pulses, then $\sigma_{P_1} = \sigma_{P_2} = (N_1 + N_2)^{-1/2}$ and $\sigma_{P_c} = (N_1 + N_2)^{-1/2}$. The statistical error of the circular polarization for stars of $9^m$ with exposure time 10 sec is usually 0.05-0.07%.

Errors in each point of the cuts were calculated in the following manner. The random errors of the degree of circular polarization in each measured area of the coma were determined by the statistics of recorded photons from the comet summarized over corresponding areas in all cuts and from the sky background. The comet was rather bright, so that in different nights and for different cuts the signal-to-noise (S/N) ratio varied from 4-20 at the ends of the cuts, where the coma was rather faint, to 12-100 in the photometric center of the coma. Consequently, the random error was mainly determined by the statistics of the counts from the comet than from the sky background. On the other hand, the random errors were calculated by the internal dispersion of the measured degree of circular polarization in the corresponding areas of the coma of the individual cuts. The larger of these two errors was adopted as an accuracy measure of the weighted mean values of circular polarization for each area in every cut. The errors of the night-averaged degree of circular polarization were determined using the variance of individual values. The typical errors in the degree of circular polarization (see Table 2) varied from 0.05 to 0.1%.

*2.4.2. Systematic errors*

The averaged instrumental circular polarization was equal to 0.066±0.008% and was carefully subtracted from the cometary data. A similar result (not exceeding 0.1%) was obtained at the measurements of the circular polarization for the unpolarized light in laboratory tests (Shakhovskoy et al., 2001). Thus, the systematic error introduced by instrumental polarization was significantly less than the random errors of the degree of circular polarization in each measured area of the coma.

To estimate the systematic errors of measurements caused by the conversion of linear polarization into circular polarization, the results of laboratory tests were used (Shakhovskoy et al., 2001). For the light with a 100% circular polarization, the efficiency of the instrument was close to 100%. For the light with 100% linear polarization, the spurious circular polarization never exceeded the limits −0.5÷+0.5%, depending on the position angle of the linear polarization. Thus, we can conclude that the linear polarization of the comet, which was within the range from 17 to 24% during our observations (Kiselev et al., 2001), could produce the maximal spurious circular polarization on the order of several hundredths of a percent, which is within the limits of the observational errors. Comet S4 was dusty, therefore the degree of linear polarization and the position angle of polarization plane should not show noticeable changes over the coma since this is typical for dusty comets (Dollfus and Suchail, 1987; Kiselev, 2003; Kiselev and Rosenbush, 2004). In such a case, the spurious circular polarization should have the same sign and value over the coma during one night. Our results (see below Figs. 1-2) show systematic variations of both the value and sign of circular polarization along the cuts. This gives us a good reason to believe that the observed variations of the circular polarization over the coma are real.

*2.4.3. Problems connected with using the wide-band filter and large diaphragm*

In addition to the continuum flux from the dust, the wide-band R filter passes emissions by molecular gases. From linear polarization measurements it is well known that the presence of gas emission in the passband of the filter reduces the measured polarization. A dilution of the linear polarization in continuum due to the molecular emission contribution is of particular importance for so-called gas-rich comets (Jockers et al., 2005). Comet S4 belongs to the class of dusty comets (Farnham et al., 2001). According to Kiselev (2003), the maximum reduction of the dust linear polarization for dusty comets does not exceed 0.08 of the maximum linear polarization, i.e. for comet S4, whose maximum polarization was equal to ~24%, the reduction may reach 2%. Under the assumption that this proportion holds for CP, we can conclude that the maximum reduction of CP in comet S4 would only be several hundredth of a percent.

The use of a comparatively large diaphragm can also lead to reduction of the degree of circular polarization. It is caused by averaging of CP with different signs within local areas of the coma. For example, if the circular polarization was caused by the grain alignment due to interaction with the radial outflow of the gas, it should be expected to be minimal when the comet is centered in the diaphragm, since the distribution is isotropic in the field accepted by the diaphragm (this may be a reason why CP measured in the near-nuclear area of the coma is usually small). However, a real comet is not an isotropic medium. Actually, different size distribution of dust particles is observed at the solar and tail part of the coma. Besides, the area of coma, where the interaction of the dust particles with the radial outflowing gas occurs, is about 20 radii out of the comet nucleus that is approximately 50 km for comet S4. In outer parts of the coma and in the tail, the CP may be produced due to the alignment of grains via other mechanisms (see section 4.2). In this case, the CP distribution over the coma, most likely, is asymmetric. Therefore it is difficult to estimate from our observational data how large CP, averaged over the diaphragm, can be in the case when the comet just enters in the diaphragm. However, in this case an effect of averaging out of CP due to the radial outflow of gas would be minimal. The diaphragm averaging due to the proper motion of the comet during a single exposure should also be small since the displacement of the comet was only ~270÷500 km (see Table 1) that is less than 0.1 of the diameter of diaphragm used. Thus, the large diaphragm as well as wide-band filter used in the observations provided a low value of the resulting CP of dust in comet S4.

## 3. Analysis of observational data

*3.1. Spatial distribution of the circular polarization*

The measured circular polarization in comet S4 was found to be at a rather high level, reaching, on average, 0.8%. We investigated its spatial variations along the cuts through the coma and nucleus in the solar and tail directions up to the distances of 6000 km. The results obtained for each date are summarized in Figs. 1 and 2.

    [Fig. 1]           [Fig. 2]

It can be clearly seen that CP shows systematic, not chaotic, variations over the coma in each night. Some measurements show rather smooth increase or decrease of CP with the distance from the nucleus, some others show more complex behavior. Although the value of CP sometimes reached 1%, the degree of circular polarization at the cometary nucleus was almost always close to zero. The left-handed as well as right-handed polarization was observed over the coma, but left circularly polarized light was systematically observed in the sunward part of the coma. On June 28, when the cut was made outside the nucleus at the tailward side of the coma, only right-handed polarization (on average 0.58±0.15%) was observed along the cut (Fig. 1). On the next three days, from June 29 to July 02, CP showed very similar behavior. Later, on July 18-19, a large outburst occurred, and the ultimate breakup of the comet happened on 19-20 July (Farnham et al., 2001; Bockelee-Morvan et al., 2001). According to Weaver et al. (2001), the complete disintegration peaked on 22.8±0.2 July. Apparently, this had an effect on the behavior of the CP. Actually, changes in the circular polarization over the coma on July 21 and 22 are very different. Immediately after the complete disintegration of the nucleus, polarization in both directions was mainly left-handed and essentially non-zero: its averaged value was $P_c = 0.41 \pm 0.07$ % that was significantly greater than $3\sigma$ (Fig. 2, third panel). However, the next day, on July 22, the trend changed to a smooth increase from negative to positive values (Fig.2, bottom panel).

Since the errors of the measurements are smaller than the variations of CP over the coma, we can conclude that these variations are statistically significant and, thus, real. Although the criterion of systematic behavior of the measured CP cannot be applied to estimate the validity of the measurements of CP for such an active comet as comet S4, there is a resemblance between the data obtained before the disruption at the end of June and at the beginning of July (see Fig. 1, 2-4 panels, and Fig. 2, 1-2 panels). This resemblance can be considered as one more evidence that the obtained data are realistic.

*3.2. Phase angle dependence*

During our observations the phase angle of the comet varied from 61 to 122°. This allowed us to study variations of the night-averaged circular polarization of the comet with phase angle. We did not find a monotonous change of this polarization with phase angle. There were significant variations in circular polarization, which followed the activity of the comet, i.e. coincided in time with the outbursts caused by the fragmentation of the nucleus.

Farnham et al. (2001) indicated that the coma in the sunward direction was essentially unaffected by the outbursts, maintaining the same basic shape and brightness throughout. Taking into account this fact, we constructed a phase-angle dependence of the circular polarization for the sunward part of the coma and for the tail part separately (Fig. 3). It turned out that the averaged degree of circular polarization along each cut in the sunward direction has clear phase-angle dependence: the absolute value of the circular polarization increases with phase angle, as seen in Fig. 3a. A trend of the averaged degree of polarization along each cut at the tail side (Fig. 3b) differs from that at the sunward side considerably.

**[Fig. 3]**

Using all available measurements of circular polarization in comets Halley, Hale-Bopp, and S4 we constructed the composite phase dependence of circular polarization (Fig. 4), which indicates an increase in absolute value of the circular polarization with the phase angle. Dollfus and Suchail (1987) also found that the degree of circular polarization of comet Halley showed a marginally significant increase with the phase angle.

**[Fig. 4]**

*3.3. Temporal variations in the circular polarization*

Previous investigations have shown that non-steady processes in comets, notably, brightness outbursts or sudden increases of dust outflow from the nucleus, can change the linear polarization (Kiselev and Chernova, 1979; Dollfus, 1989; Rosenbush et al., 1994). Comet S4 was an actively outgassing, highly fluctuating object, and during its approach to the Sun displayed sporadic outbursts due to partial and then complete disintegration. As a result of the comet disintegration, the internal parts of the nucleus were exposed and fresh dust particles appeared in the cometary atmosphere increasing the dust and gas production rates. Moreover, the dust particles, lifted from the nucleus, were continuously losing their volatiles and fragmenting. In consequence of this, there were noticeable changes in the dust properties (sizes, composition, and structure) with the distance from the nucleus (Bonev et al., 2002). These changes mainly affected the tailward side (Bonev et al., 2002), whereas, as it was pointed out above, the coma in the sunward direction was essentially unaffected by the outbursts. This may explain why Fig. 3b looks more erratic than Fig. 3a.

We compare the temporal variations in the comet CP with the variations in other comet characteristics: visual magnitude, which is sensitive to the dust and gas production rates, water production rate, and linear polarization (Fig. 5). The night-averaged visual magnitude of comet S4 was compiled from ICQ (2000) and reduced to the geocentric distance of 1 AU. Water production rate for our observational period was taken from Mäkinen et al. (2001), Bockelée-Morvan et al. (2001), and Farnham et al. (2001). We also show the linear polarization data from Kiselev et al. (2001). In order to study temporal variations of the linear polarization, the data were corrected for phase-angle dependence. For this, the polarization data were approximated by a polynomial and the bottom panel of Fig. 5 presents deviations ($\Delta P_{lin}$) of the observed linear polarization from the best-fit curve.

[Fig. 5]

The analysis of the temporal changes in comet brightness (Fig. 5) revealed several large outbursts, which reached their maxima around July 6-8, 16-18, and 22-23 with smaller ones in between. These outbursts correlate with increasing water production (Mäkinen et al., 2001). Between July 18 and 23 the visual brightness of comet increased up to $0.8^m$. During this period the $H_2O$ production rate increased by a factor of ~10 (Bockelee-Morvan et al., 2001). The ultimate breakup of comet S4 started on July 18 (Bockelee-Morvan et al., 2001). The maximum of the fragmentation event activity of the comet occurred on 22.8 July (Weaver et al., 2001) that is in a good agreement with the maximum of brightness. Thus, the comparison of the total visual magnitude with the water production rate shows that there is some correlation between the time of fragmentation and the increase in brightness and gas and dust production rate of the comet.

The nucleus disintegration caused changes in the phase-angle dependence of dust polarization. Degree of linear polarization sharply increased after complete disintegration (Kiselev et al., 2001; Hadamcik and Levasseur-Regourd, 2003). It is unlikely to expect a strong correlation between the variability of cometary brightness and the degree of polarization because the visual magnitude of a comet strongly depends on the $C_2$ emission, not only on the strength of the continuum. Nevertheless, we notice a consistency between the variations of linear polarization and the comet activity. The measurements of the circular polarization also support this view: in Fig.5 one can see that the CP is strongly correlated with the linear polarization.

## 4. Discussion

Most likely, circular polarization in comets is a result of a complex of phenomena, which take place in the cometary atmosphere. In this paper we do not offer a comprehensive interpretation of the data, but instead endeavor to confirm that there are mechanisms that can operate in comets and generate the results consistent with our data.

The mechanisms that may produce circular polarization in comets were considered by Bandermann and Kemp (1973), Dolginov and Mytrophanov (1975, 1976), Beskrovnaja et al. (1987), Dolginov et al. (1995). Summarizing these studies, one can conclude that circular polarization in comets may appear as a result of:

- multiple scattering in an anisotropic medium;
- scattering by aligned nonspherical particles;
- scattering by optically active (chiral) particles.

The following consideration reveals the fundamental reasons that are responsible for the formation of CP by the above mechanisms; then we consider each of the mechanisms in more detail.

In order to understand how circular polarization can be produced by cometary dust, let us imagine that sunlight interacts with some ensemble of dust particles. In the general case the Stokes vector of the incident light $(I_0, Q_0, U_0, V_0)$ gets transformed into the Stokes vector of the scattered light $(I, Q, U, V)$ through the scattering (Mueller) matrix $\mathbf{S}$, i.e.

$$\begin{vmatrix} I \\ Q \\ U \\ V \end{vmatrix} = \mathbf{S} \times \begin{vmatrix} I_0 \\ Q_0 \\ U_0 \\ V_0 \end{vmatrix} = \begin{vmatrix} S_{11} & S_{12} & S_{13} & S_{14} \\ S_{21} & S_{22} & S_{23} & S_{24} \\ S_{31} & S_{32} & S_{33} & S_{34} \\ S_{41} & S_{42} & S_{43} & S_{44} \end{vmatrix} \times \begin{vmatrix} I_0 \\ Q_0 \\ U_0 \\ V_0 \end{vmatrix}. \tag{5}$$

If an ensemble of particles is illuminated by unpolarized sunlight, the Stokes vector of the incident light is equal to $(I,0,0,0)$. It is clear that in this case after a single scattering CP will be non-zero only if matrix element $S_{41}$ is non-zero.

As was shown by van de Hulst (1981, section 5.22), and more fully in (Hovenier and van der Mee, 1996; Hovenier and Mackowski, 1998), if an ensemble of particles consists of equal number of particles and their mirror images, then the scattering matrix becomes

$$\begin{vmatrix} S_{11} & S_{12} & 0 & 0 \\ S_{21} & S_{22} & 0 & 0 \\ 0 & 0 & S_{33} & S_{34} \\ 0 & 0 & S_{43} & S_{44} \end{vmatrix}. \tag{6}$$

This ensemble of particles cannot produce CP since $S_{41} = 0$. Spheres (which coincide with their mirror particles) or randomly-oriented rotationally symmetric particles (e.g. spheroids) have a scattering matrix of this type, and such ensembles do not produce CP. However, if the number of particles and their mirror particles is not equal, we get van de Hulst's case (5.22.1):

$$\begin{vmatrix} S_{11} & S_{12} & S_{13} & S_{14} \\ S_{21} & S_{22} & S_{23} & S_{24} \\ -S_{13} & -S_{23} & S_{33} & S_{34} \\ S_{41} & S_{42} & -S_{34} & S_{44} \end{vmatrix}. \tag{7}$$

In this case $S_{41}$ is not zero and CP can appear as the light is scattered by this ensemble of particles. Such a situation appears at any violation of mirror symmetry in the ensemble. It can be intrinsic mirror-asymmetry of particles (chirality, see section 4.3) or a case of violation of their random distribution, e.g. when elongated particles are not randomly oriented.

Circular polarization can be also produced through multiple scattering. In this case the first scattering, as seen from matrix (6), produces the Stokes vector $(S_{11}, S_{12}, 0, 0)$. If the second or any high-order scattering happens in the scattering plane that is different from the scattering plane of the first scattering, then the Stokes vector gets non-zero third Stokes parameter due to implementation of the rotational matrix (see, e.g., Mishchenko et al., 2002)

$$\begin{vmatrix} 1 & 0 & 0 & 0 \\ 0 & \cos 2\eta & -\sin 2\eta & 0 \\ 0 & \sin 2\eta & \cos 2\eta & 0 \\ 0 & 0 & 0 & 0 \end{vmatrix}, \qquad (8)$$

where $\eta$ is the angle between the scattering planes of the first and second scattering. Then after the following scattering, i.e. one more multiplication by matrix (6), the fourth Stokes parameter becomes non-zero and the light becomes circularly polarized.

*4.1. Multiple scattering*

Multiple scattering as described above is a mechanism often postulated as a source of circular polarization in a variety of dust-containing environments. For multiple scattering to produce circular polarization, neither special shape nor special composition of particles is necessary. However, there are the following two requirements for the mechanism to operate:

- The spatial distribution of the dust particles must be asymmetric. Otherwise, the amount of the produced left-handed CP will be equal to the amount of right-handed CP and the total circular polarization from the ensemble of particles will be zero (see, e.g., Kemp, 1974; Degtyarev and Kolokolova, 1992; Whitney and Wolff, 2002).

- A significant optical depth of the dust is necessary so that the multiple scattering can occur.

The first condition (asymmetry), most likely, is valid for comet atmospheres, which are characterized by non-spherical shape and the presence of a variety of structural features that lack symmetry of any sort. At small heliocentric distances, comets often show dust jets and sunward fans, apparently due to non-uniform surface activity of the nucleus. The morphology of the dust jets is influenced by radiation pressure on the dust, as well as by nucleus rotation that makes the jets asymmetric with respect to the center of brightness.

According to van de Hulst (1981, section 1.22), multiple scattering becomes important at optical thicknesses $\tau > 0.3$. However, significant optical thicknesses of the dust, is doubtful for most comets (see, e.g., Jewitt, 1989; Rosenbush et al., 1997). Among the comets for which CP has been observed, namely Halley, Hale-Bopp, and S4, the largest $\tau$ was apparently for comet Hale-Bopp. As it was shown from the observations of a stellar occultation, the coma of comet Hale-Bopp had $\tau > 1$ within the distance 100 km from the nucleus at ~3 AU (Fernandez et al., 1999, Weaver and Lamy, 1999). Similar observations for the same comet by Rosenbush et al. (1997) at 1.39 AU showed $\tau \approx 0.3$ at the distance $10^5$ km from the nucleus. *In-situ* observations of jets in comet Halley (Keller et al., 1987) showed the dust optical thickness 0.28 at a distance of 6500 km, while Dollfus and Suchail (1987) visually estimated $\tau \approx 0.92$ at the distance of about 500 km. Although a non-zero, $\leq 0.3\%$, circular polarization was measured for the near-nucleus region of comet Hale-Bopp (Rosenbush et al., 1997; Manset and Bastien, 2000) as well as for comet Halley, in both cases the data on circular polarization and optical thicknesses were obtained for different areas of the comae, and this does not allow us to estimate the influence of multiple scattering on formation of the circular polarization. As far as we know, there are no direct measurements of the dust optical thickness for comet S4. However, since the nucleus radius of this comet was small, about 0.44 km (Farnham et al., 2001), it is difficult to expect a large optical thickness there. Thus, we do not completely rule out multiple scattering as a source of circular polarization in bright dust-rich comets, but it is unlikely that it worked in the case of comet S4.

*4.2. Scattering by aligned non-spherical particles*

The most discussed mechanism for producing CP in comets is scattering on aligned non-spherical particles[1] (see Dolginov et al., 1995; Lazarian, 2003, and references therein). The existence of oriented particles was directly discovered in observations of a star occulted by the coma of C/1990 K1 (Levy) as the starlight transmitted through the coma showed changes in the value of linear polarization and position of its plane (Rosenbush et al., 1994). Since at an occultation the star radiation was forward-scattered, this transformation of polarization pointed to the alignment of dust grains in the atmosphere of the comet. Polarimetric observations of a stellar occultation by comet Hale-Bopp confirmed the results (Rosenbush et al., 1997). The fact that cometary grains are non-spherical is widely recognized now (see, e.g. Kolokolova et al., 2004a). Thus, they are capable of alignment. A strong constraint on the mechanism that can align comet dust is that it should be rapid. Indeed, the significant value of CP was observed at distances of 500-1000 km from the nucleus of comet S4. Dust particles with velocities of hundreds meters per second (Farnham et al., 2001) need $\sim 2.5 - 3 \cdot 10^3$ s to reach these distances.

The most common textbook explanation of grain alignment is related to the paramagnetic relaxation first described by Davis and Greenstein (1951). However, the corresponding relaxation rates are excessively low to be important for comets (although it can play some role in additional to more efficient mechanisms, see below).

Two other mechanisms that may align dust particles in cometary atmospheres are relative gas-grain motion (henceforth mechanical alignment) and radiative torques. Outflowing gas could play an important role in grain alignment near the comet nucleus ($\sim 10$-$20$ $R_{nucleus}$). In outer parts of the coma and in the tail, the alignment via radiative torques and interaction with solar wind could be important. For the comet environment the most important are mechanical alignment of grains rotating thermally (Gold, 1952; Purcell, 1979; Dolginov and Mytrophanov, 1976; Lazarian, 1994, 1997a,1997b; Roberge et al., 1995) and suprathermally (Lazarian, 1995; Lazarian et al., 1996; Lazarian and Draine, 1997; Lazarian and Efroimsky, 1999) and radiative torque alignment (Dolginov, 1972; Dolginov and Mytrophanov, 1976; Draine and Weingartner, 1996, 1997; Weingartner and Draine, 2003; Abbas et al., 2004; Cho and Lazarian, 2005). For the review of the mechanisms of grain alignment see Lazarian ( 2003).

---

[1] Although birefringent particles aligned in such a way that their optical axes have some predominant orientation can also produce circular polarization, we are not aware of any natural phenomena that can align cometary particles this way. So, in this paper we will not consider birefringent aligned particles as a source of cometary circular polarization.

Consider mechanical alignment first. Potentially mechanical alignment can align grains on time scales less than the gaseous damping time (the latter is time required by the radiative torques). In this situation we may see mechanically aligned grains at distances less than $10^3$ km. To be efficient, mechanical grain alignment requires grain-gas velocities larger than the thermal velocity of gas molecules (see discussion in Lazarian (1997a)), and it happens differently for grains rotating thermally and much faster than thermally (suprathermally). The evaporation of the molecules from a grain will induce torque that preserves the same direction in the frame of the grain (Purcell, 1979) and can make grains suprathermal. Dealing with interstellar conditions, Lazarian and Draine (1999b) demonstrated that grains perform frequent thermal flipping, which results in Purcell's torques are being averaged out. This leaves most of the interstellar grains rotating thermally ("thermally trapped") unless they are spun up by radiative torques. In comets, if we consider alignment over times that are shorter than the Barnett relaxation time, thermal trapping is not important. The calculations of the degree of alignment for the imperfect internal alignment (i.e. the alignment of grain axis of maximal moment of inertia and angular momentum ) in Lazarian (1997a) indicate that the degrees of alignment up to 10% are achievable. Such an alignment would be characterized by grains having their longer axes directed preferentially along the relative grain-gas flow due to the tendency of the grain to minimize its cross-section in the gas flow.

Alignment due to radiative torques was first discussed by Dolginov and Mytrophanov (1976). The essence of the radiative torques is simple: the difference in the extinction cross sections for photons with different signs of circular polarization results in the difference in the number of scattered left- and right-polarized photons. As a result of the recoil arising from differential extinction the irregular grain gets spun up. While Dolginov and Mytrophanov (1976) thought that special shapes might be necessary, the numerical calculations in Draine and Weingartner (1996) show that it is sufficient for grains to be irregular. The application of the radiative torque alignment idea to comets requires some additional considerations (Dolginov and Mytrophanov, 1976; Lazarian, 2003).

To evaluate the efficiency of radiative torques for comet S4, we calculated numerically radiative torque for an irregular dust particle made of astronomical silicate (Laor and Draine, 1993). Following Kimura et al. (2003), we considered comet particles as ballistic aggregates consisting of 128 monomers of size of $10^{-5}$ cm. For the calculations the DDSCAT package (Draine and Flatau, 1994) was used. The gas drag was obtained using a model of comet environment that was based on the observations by Altenhoff et al. (2002). Comet S4 had gas production equal $9\times10^4$ g/s. Since the radius of its nucleus was approximately 440 m, and the speed of the gas on the order of 1 km/s, the density of gas that was $2.5\times10^{11}$ mol/cm$^3$ near the nucleus, decreased as $\rho^{-2}$ where $\rho$ is the distance from the nucleus, reaching values on the order of $10^5$ mol/cm$^3$ at the distances where circular polarization was observed. Our calculations show that for comet environment, where gas damping and radiative torque compete, we can easily get radiative torque to dominate due to the low density of comet gas. At the above mentioned values of number density of gas molecules, the nearly complete alignment of particles by radiative torque is achieved at densities corresponding to the distances on the order 120 km from the nucleus. However, this is a lowest boundary for the alignment radius. The maximum speed the comet dust can reach is about v=1 km/s. The radiative alignment happens over the damping time which is on the order of $10^3$ s for the grains considered.

Therefore we expect alignment can be attained within ~$10^3$ km from the nucleus. This value is comparable with the distance from the nucleus where CP was observed.

The efficiency of radiative torques increases fast with grain size (Cho and Lazarian, 2005). However, it takes longer for larger grains to be aligned since their alignment time depends on the gas damping time, which is longer for larger particles. This may compensate for the enhanced efficiency of radiative torques and we might expect larger grains to be aligned further away from the nucleus.

For quantitative studies of grain alignment by radiative torques in comets atmospheres additional effects should be taken into account. For instance, internal grain wobbling can be induced by thermal fluctuations within the grain material (Lazarian, 1994). The characteristic time of this wobbling is equal to the time-scale of the relaxation due to internal dissipation of energy within a wobbling grain (Lazarian and Draine, 1997; Lazarian and Roberge, 1997). Purcell (1979) identified the so-called Barnett relaxation as the dominant process of internal relaxation. This type of relaxation is related to the variations of the rotation-induced magnetization as the grain wobbles and it has a characteristic time scale of $10^4 T_{grain}^{-1} \omega_6^{-2}$ s, where the grain angular velocity $\omega$ is normalized over $10^6$ s$^{-1}$ and the grain temperature is taken to be 300 K for the estimate. Further research in the field identified nuclear relaxation (Lazarian and Draine, 1999a) as the dominant internal relaxation process for grains rotating slower than $10^{-5}$ s$^{-1}$. As the result, the alignment of angular momentum is expected to be only partial at the distances from the nucleus smaller than $10^3$ km. However, the wobbling should not interfere with the alignment as soon as grain gets velocities substantially larger than the thermal Brownian one. Therefore for distances from the comet nucleus larger than $10^3$ km we do not expect to observe much interference due to grain wobbling.

Lazarian (2003) estimated that radiative torques should dominate grain alignment for a typical comet environment. Our considerations above also agree with this conclusion. However, the joint action of two mechanisms may result in interesting and not yet explored transient alignment processes that can influence the CP pattern observed. Note, that if the alignment is produced only by radiative torque, then particles get aligned with long axis perpendicular to the radiation. In this case for an ensemble of random particles the average circular polarization should be equal to zero. It can be clearly seen from eq. (7) of Lazarian (2003) since in this case the triple vector product vanishes: $[e_0 \times e_1]e = [e_0 \times e_1]e_0 = 0$. However, interplay with mechanical alignment and influence, even very weak, of magnetic of electrical field, which can tilt the particles from the position perpendicular to the radiation, immediately result in non-zero circular polarization.

In interstellar medium gas-grain alignment always happens with respect to magnetic field. Indeed, this happens if the time-scale of grain alignment is longer than the period of grain precession in the external magnetic field. The latter for an ordinary paramagnetic grain is approximately $4 \times 10^4 B_{-4}^{-1} a_{-5}^2$ s, where the grain size, $a$, is normalized over a reference $10^{-5}$ cm size and the magnetic field, B, is normalized by $10^{-4}$ G. However, the Larmor precession will be faster by the ratio of the grain paramagnetic susceptibilities if grains are superparamagnetic, ferrimagnetic or ferromagnetic (see Morrish, 1980; Roberge and Lazarian, 1999). This ratio may be larger than $10^3$ decreasing the Larmor period to seconds. However, the Larmor precession itself does not guarantee the particular sense of alignment, e.g. alignment may happen perpendicular as well as parallel to the magnetic field. It is possible to show that for comet grains the Larmor precession is only important if grains are strongly magnetic or at the distances larger than $10^3$ km from the comet nucleus.

If the Larmor precession is slow compared to the grain alignment the alignment that happens will be with respect to the flow of radiation or photons. Radiative torques in the absence of Larmor precession tends to align grains with long axes perpendicular to the direction the flow (see Lazarian, 2003). If grains are strongly magnetic (e.g. have ferromagnetic inclusions) their Larmor rotation is fast and the alignment happens with respect to magnetic field. Similarly, grains far from the comet nucleus interact with tenuous gas. As the radiative torque alignment happens of the time scale on the order of the gaseous damping time which is $10^3 n_{12} a_{-5} T_{100}^{-1/2}$ s, where the density is normalized over $10^{12}$ molecules. Therefore at several times $10^3$ km we expect to see a change of alignment to reflect the magnetic field direction. While we shall evaluate the precise radius within more elaborate models of comet atmospheres, we note that by itself the change of the alignment direction can be a cause of the emergence of circular polarization.

The CP data poses serious questions to the current grain alignment theory that was developed mostly in order to explain diffuse interstellar and molecular cloud data. If grain alignment happens in relation to the direction Sun, it is possible to show that the expected degree CP is zero. This opens several possibilities. First of all, the alignment be mechanical and happen in respect to the local gas outflow. Then, the radiative alignment may be modified in the presence of the outflow and the direction of the alignment may get different from the direction from the direction towards the Sun. Finally, external fields may affect the alignment direction. While in the context of interstellar medium magnetic fields are dominant, they may not penetrate too close to the comet nucleus to provide sufficiently fast precession even for grains with magnetic inclusions. However, magnetic field may contribute in the circular polarization out of the magnetic cavity, i.e. at distances more than 5000 km from the nucleus (Neubauer, 1987; Liu, 1999).

In the case of comet, electric field may be a promising means for grain alignment. D. Gough (private communication) informed one of us (AL) that dust alignment is feasible in the electric field of the Earth atmosphere due to the action of the electric field on grains with dipole moments. Hoang and Lazarian (in preparation) studied a feasibility of the alignment of dust grains with dipole moment in much weaker electric field, but in the presence of radiative torques. They found that the electric field can play a role similar to one usually prescribed to magnetic field in the process of alignment by radiative torques. Namely, dipoles process around the electric field direction and it gets the new axis of alignment that does not coincide with the direction towards the Sun.

Note, that without going into the relevant alignment physics the results from the light scattering by aligned non-spherical particles was repeatedly calculated for a variety of applications (see, e.g., Beskrovnaja et al., 1987; Li and Greenberg, 1997; Gredhill and McCall, 2000). One can see from these papers that varying size, composition, and shape of the particles it is possible to obtain a broad range of CP values, including those that were observed for comet S4. For example, the phase-angle dependence for aligned Rayleigh particles calculated by Beskrovnaja et al. (1987), shown by dashed lines in Fig. 4, demonstrates a tendency similar to the observed one, namely, circular polarization linearly increases with the phase angle, although the theoretical trend is much steeper. A more careful analysis based on calculations of the light scattering by realistic non-spherical particles may provide a better fit.

*4.3. Scattering by particles made of optically active (chiral) materials*

As we showed in the beginning of this section, circular polarization indicates a violation of symmetry in the medium. This violation of symmetry can be an intrinsic property of the dust particles. For example, the particles can be chiral, i.e. not identical to their mirror images. Chirality in nature is most known as a property of complex organic molecules, existing in two forms: L (left-handed) and D (right-handed), which are identical in all respects except that they are "twisted" in opposite directions, looking like mirror images of each other.

Non-living systems normally contain equal numbers of the L and D enantiomers of such molecules. But for terrestrial biomolecules there are only L-amino acids and D-sugars. This property is called "homochirality". As a result of homochirality or any other enantiomeric excess, the media that contain biomolecules possess optical activity, a property that is also called "circular birefringence" and is a consequence of different refractive indices for polarization of different handedness. In an optically active medium, light characterized by CP of different signs has different speed. This separates left- and right-handed polarized waves and rotates the plane of linearly-polarized light (see for details Bohren and Huffman (1983), section 8.3). The effects are especially strong for organic molecules since they are more than optically active, that is, they possess circular dichroism, i.e. such substances have different absorption for left- and right-handed CP (see, e.g., Wolstencroft et al., 2004).

For a long time it was believed that homochirality, or asymmetry in the number of L and D biomolecules, is of Earth origin and relates to the birefringence of some Earth minerals. But then L-enantiomeric excess was found in amino-acids from the Murchison and Murrey meteorites (Pizzarello and Cooper, 2001), suggesting an origin in the pre-solar nebula. This idea was confirmed when high CP was measured in star-forming regions (see review in Hough et al., 2001). The illumination of cosmic organics by circularly polarized light in protoplanetary nebulae is an increasingly popular idea regarding the origin of homochirality. In that case enantiomeric excess in organics should be found not only in meteorites but also in other primitive bodies, including comets. The search for such an excess in cometary organics is the goal of one of the projects of the Rosetta mission (Thiemann and Meierhenrich, 2001). Chiral organics can also be detected remotely, by studying CP in comets, since chiral organics with an enantiomeric excess are optically active, i.e. have different refractive indices for left- and right- handed CP.

We study the effect of chirality on light-scattering properties of cometary dust considering particles that possess optical activity. We use the theoretical solution for an optically active sphere (Bohren and Huffman, 1983). Even though nobody now believes that comet dust can be represented by spherical particles, we decided to try this model as a first approximation. To make the calculations more realistic, we consider the particles with the power-law size distribution measured *in situ* for the dust in comet Halley (McDonnell et al., 1987). We use the values of the refractive indices and specific optical rotation typical for the amino acids discovered in the Murchison meteorite (Cronin and Pizzarello, 1986). According to Mason (1982), the specific rotation angle for them is about 100°, which corresponds to the difference in the refractive index for the left and right circular polarization $3\times10^{-6}$. In our calculations only 10% of the material is supposed to be optically active (chiral). Under these conditions we cannot obtain values of circular polarization larger than 0.005%. However, taking into consideration the circular dichroism of organic molecules (i.e. different absorption for left and right circularly polarized light) significantly increases calculated circular polarization. Even though circular dichroism is usually very small, making the difference in the imaginary part of the refractive index on the order of $1\times10^{-8}$ (Schreier et al., 1995), here it appeared to be sufficient to produce circular polarization that has the observed phase-angle trend and reaches 0.15% at the phase angle 120°.

Even though in this case the calculated values are smaller than the observed ones, there are still many mechanisms that can increase the CP produced by light scattered on chiral molecules. For example, the CP is significantly larger for specific sizes of the particles, usually in the range 1-5 μm; the considered broad size distribution (from 0.01 to 100 μm) is one of the factors that decrease the values of the CP. It should be pointed out that a change in particle size influences CP significantly and can even change its sign that may explain different signs of CP in different parts of coma. Also, CP may be larger for the materials at low temperatures: certain experiments (see, e.g. Bai et al., 2002) show that the angle of optical rotation gets 3 times larger as the temperature drops from 260 to 230 K. Finally, the realistic shape and structure of cometary grains can be a crucial factor in the genesis of the circular polarization observed. For example, the most realistic model of comet dust, the aggregate model (see, e.g. Kolokolova et al., 2004b), is characterized by optical interaction between the constituent particles that provides an effect similar to the effect of the multiple scattering discussed in Section 4.1 but stronger due to proximity of the interacting particles. Thus, the small CP produced by a single optically-active (e.g. containing chiral organics) particle should increase significantly if such particles are arranged in an aggregate. This, together with the spectroscopically detected enriched organic composition of comet S4 (Mumma et al., 2001), allows us to reasonably speculate that CP may indicate the presence of prebiotic organics in this comet.

**5. Conclusion**

We presented the results of the measurements of circular polarization in comet C/1999 S4 (LINEAR) that was exceptional in many respects, but mainly because its nucleus underwent multiple fragmentations culminating in the complete disintegration of the comet around 20 July 2000. Circular polarization of the light, scattered by the dust in this comet, was detected up to the values of 0.8%, with some trends over the coma.

We suggested several mechanisms that may be responsible for the circular polarization in comets. Among them the most promising mechanism is scattering of light on aligned non-spherical particles or on particles made of optically active (chiral) materials. The composite phase-angle dependence of CP of comets S4, Halley, and Hale-Bopp shows an increase of CP with the phase angle (by absolute value). This is in rather good agreement with the theoretical simulations for aligned Rayleigh particles, although it is not as steep. The calculations of the CP produced by spherical particles with Halley-type size distribution and containing optically active (chiral) organics show a similar phase-angle trend, although they do not reach the values of CP larger than 0.15%. However, the observed values of CP might be obtained using a more realistic, that is aggregate, model of comet dust if the aggregate constituent particles contain chiral materials.

We found the correlation between the degree of circular polarization, visual magnitudes, water production rate, and linear polarization of comet S4 in June-July 2000. For the moment our analysis does not allow us to suggest a reason for these correlations. This could be done after a systematic study of the dependence of CP on size, composition, and shape of dust particles for all possible mechanisms of CP formation. It may be that these correlations serve as a litmus paper that will help us to identify the mechanism(s) responsible for the circular polarization observed in comets. Such a study is a subject of future research.

The main purpose of this paper was to present the new observational results on circular polarization of unique comet C/1999 S4 (LINEAR). We also provide an extended discussion of the mechanisms that can produce circular polarization in comets and show that the observed phenomena are not straightforward to explain them with the present theory. We indicated the most promising directions in the theoretical study of circular polarization in comets and showed that they might shed light on mechanisms of particle alignment in comet atmosphere or/and content of complex organics in comet material. We hope this paper stimulates future theoretical and observational efforts in studying comet circular polarization.

**Acknowledgement**

We are very grateful to S. Kolesnikov for the help in the data reduction and to the anonymous referees for the suggestions which helped us to significantly improve the paper. AL acknowledges the support of the NSF grant AST 02-43156 and the NSF funded Center for Magnetic Self-Organization in Laboratory and Astrophysical Plasmas.

Table 1

Observational log of Comet C/1999 S4 (LINEAR)

| Date, UT 2000 | $r$ AU | $\Delta$ AU | $\alpha$ deg | PA deg | $PA_{cut}$ deg | $\rho$ km | $\Delta\rho$ km |
|---|---|---|---|---|---|---|---|
| June 28.975 | 0.928 | 1.066 | 60.9 | 271.0 | 37.4 | 5799 | 270 |
| 29.989 | 0.917 | 1.029 | 62.7 | 271.3 | 38.3 | 5598 | 281 |
| 30.972 | 0.907 | 0.993 | 64.5 | 271.7 | 40.0 | 5402 | 295 |
| July 02.971 | 0.888 | 0.920 | 68.4 | 272.4 | 42.2 | 5005 | 328 |
| 08.971 | 0.835 | 0.702 | 82.3 | 278.6 | 54.0 | 3818 | 501 |
| 09.996 | 0.828 | 0.666 | 85.1 | 280.7 | 56.9 | 3623 | 552 |
| 21.839 | 0.770 | 0.375 | 121.3 | 34.0 | 114.7 | 2038 | 582 |
| 22.854 | 0.768 | 0.372 | 122.1 | 47.5 | 123.6 | 2026 | 504 |

Table 2

Mean absolute value of circular polarization over the cut in comet C/1999 S4 (LINEAR)

| Date, UT 2000 | $\alpha$ deg | $P_c$ % | $\sigma_{P_c}$ % | N |
|---|---|---|---|---|
| June 28 | 60.9 | 0.788 | 0.094 | 19 |
| 29 | 62.7 | 0.480 | 0.083 | 19 |
| 30 | 64.5 | 0.526 | 0.076 | 26 |
| July 02 | 68.4 | 0.400 | 0.072 | 21 |
| 08 | 82.3 | 0.423 | 0.069 | 26 |
| 09 | 85.1 | 0.419 | 0.088 | 20 |
| 21 | 121.3 | 0.342 | 0.050 | 20 |
| 22 | 122.1 | 0.610 | 0.106 | 15 |

**Figure captions:**

Fig. 1. Variations of the degree of circular polarization along the cuts through the coma and nucleus of comet C/1999 S4 (LINEAR). The position angle of the cuts obtained on June 28-July 2 is $PA \approx 40°$. Filled circles correspond to the running average fit to the data. The error bars are standard errors of the mean value of polarization in each point of the cuts.

Fig. 2. Same as Fig.1, but for July 8, 9, 21, and 22. $PA$ is about 55° and 151° for cuts obtained on July 8-9 and July 21-22 respectively.

Fig. 3. Mean degree of circular polarization in the solar and anti-solar direction of the coma of comet C/1999 S4 (LINEAR) as a function of the phase angle. The error bars are standard errors of the mean value of polarization for the night.

Fig. 4. Composite phase-angle dependence of circular polarization for comets C/1999 S4 (LINEAR), 1P/Halley, and C/1995 O1 (Hale-Bopp). Filled circles correspond to the data for comet S4 (LINEAR); open circles and squares are the data of Manset and Bastien (2000) and Rosenbush et al. (1997) for comet Hale-Bopp; open triangles and diamonds are the data from Dollfus and Suchail (1987) and Morozhenko et al. (1987) for comet Halley. The error bars are standard errors of the mean value of polarization for the night. The solid line (1) is the linear best fit to the data. The dashed lines (2) show the results of calculations of circular polarization for aligned Rayleigh particles (Beskrovnaja et al., 1987).

Fig. 5. Comparison of the degree of circular polarization with the visual magnitudes (ICQ, 2000), water production rate (Mäkinen et al., 2001; Bockelée-Morvan et al., 2001; Farnham et al., 2001), and linear polarization of comet S4 (LINEAR)) during its final fragmentation in July 2000. The linear polarization is corrected for phase angle dependence ($\Delta P_{lin}$).

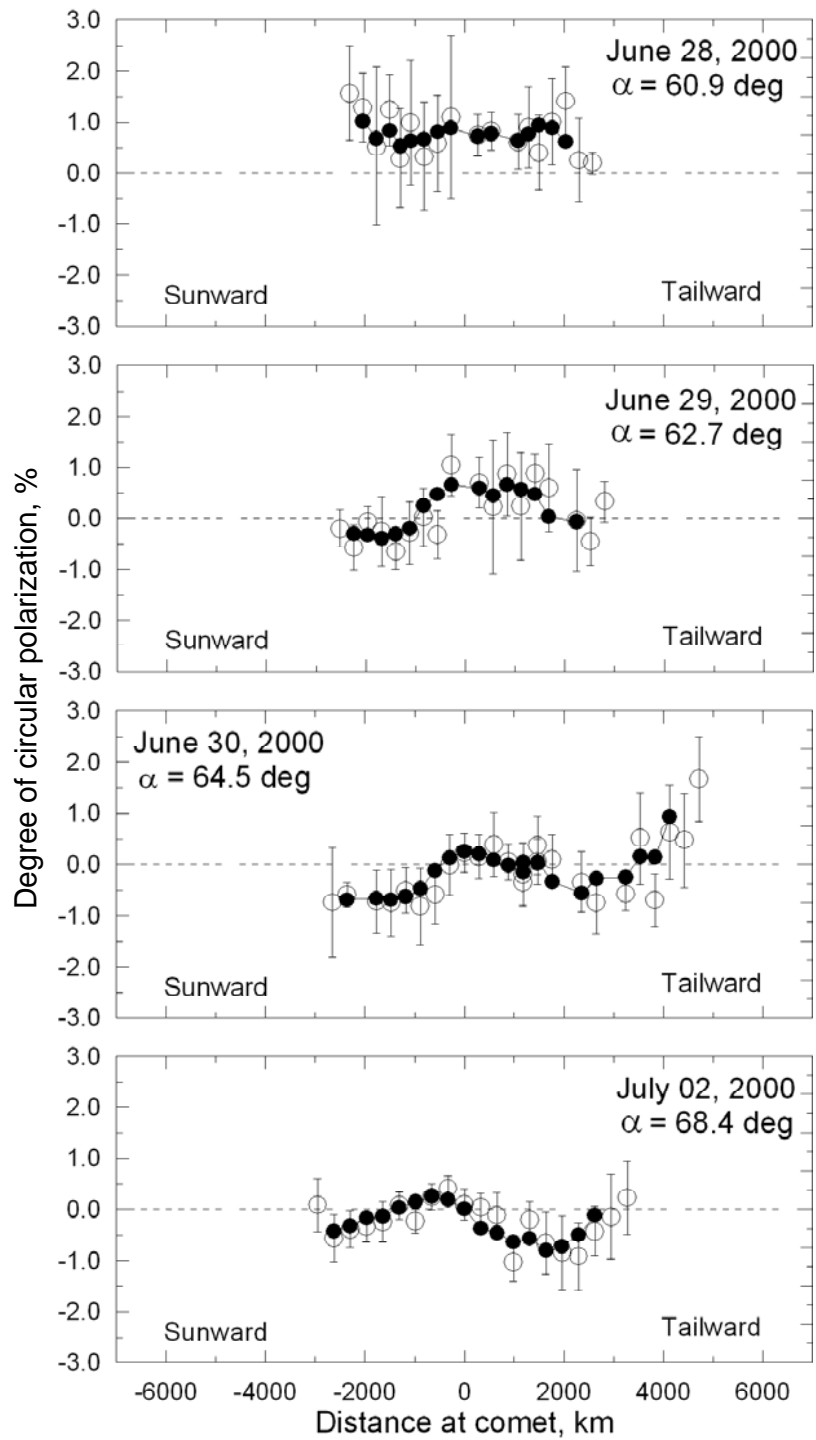

[Fig. 1]

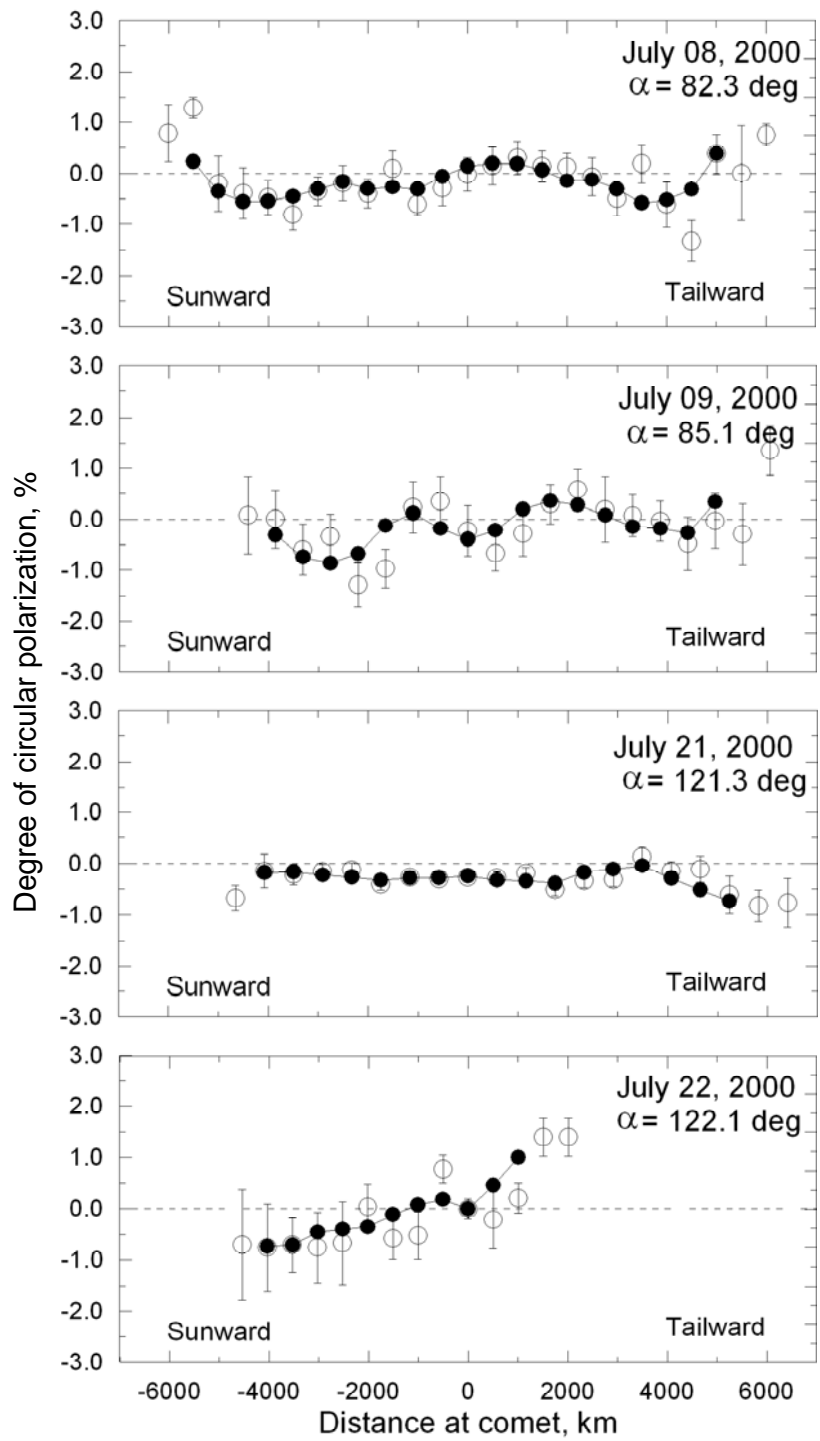

[Fig. 2]

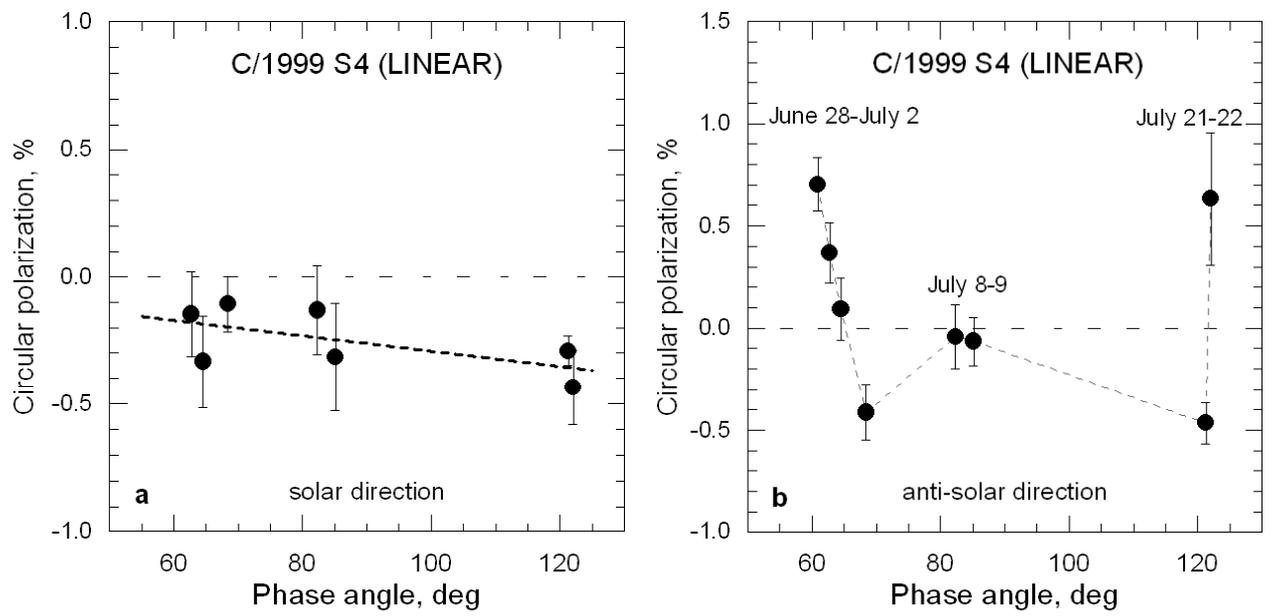

[Fig. 3]

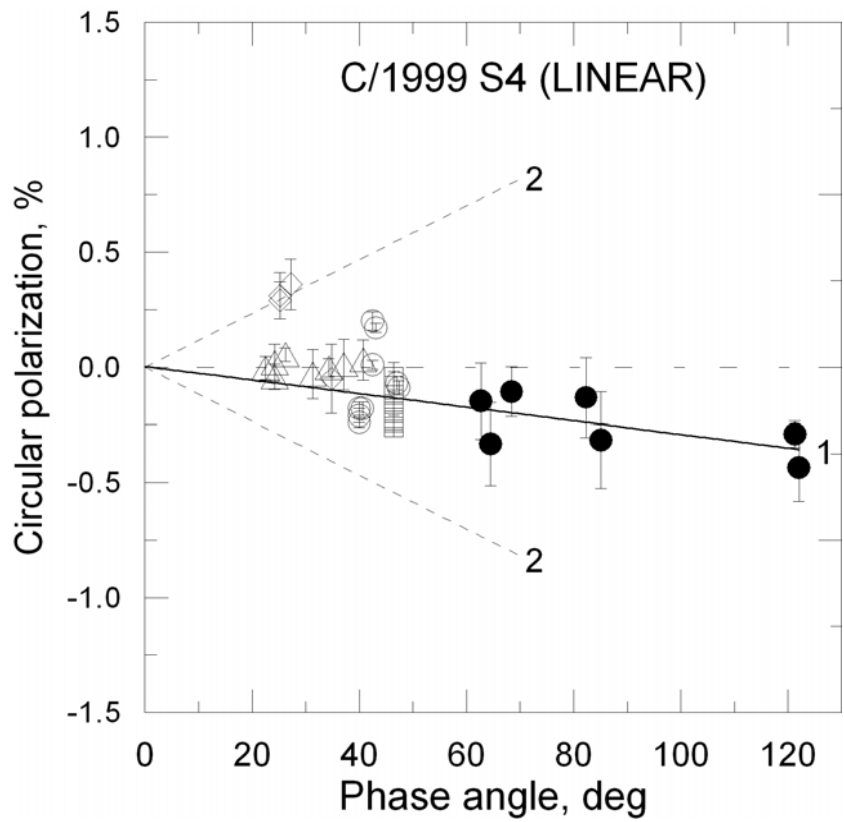

[Fig. 4]

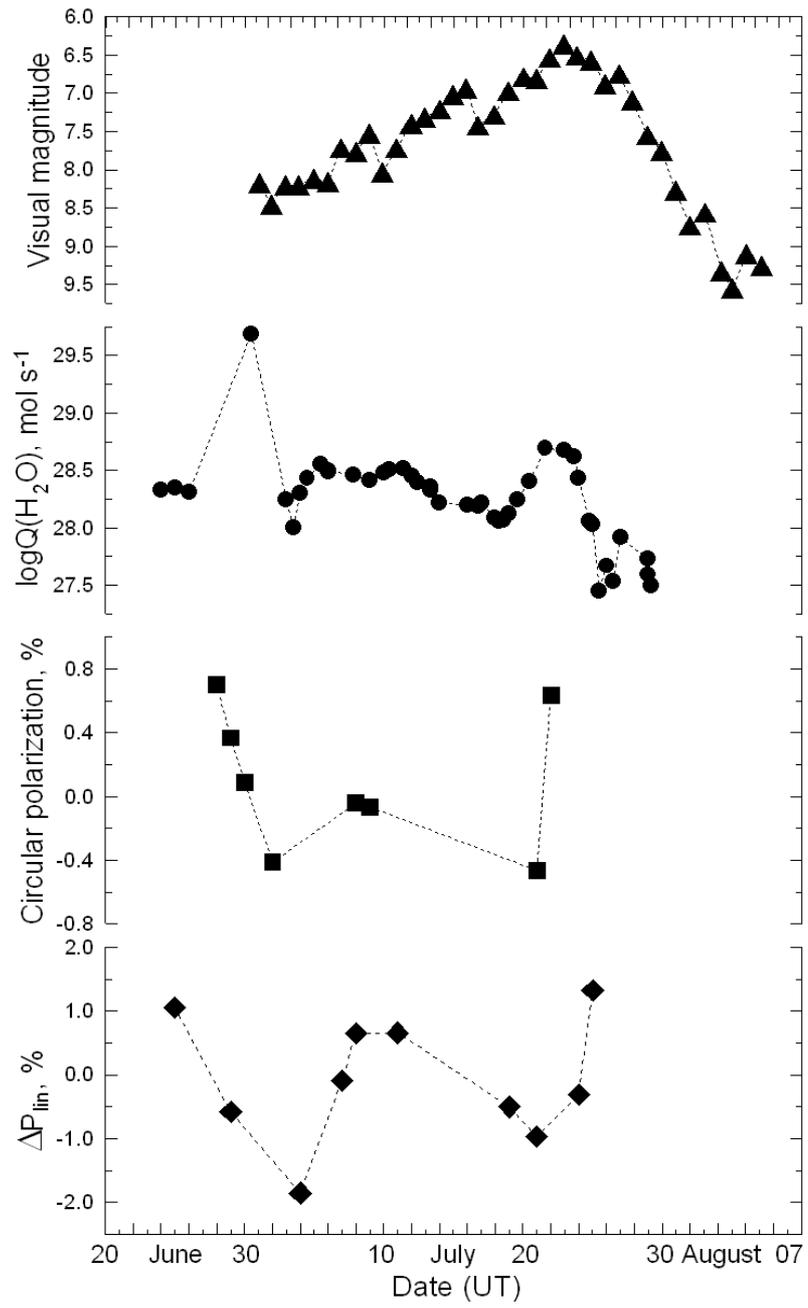

[Fig. 5]